\documentclass[aps,pra,twocolumn,groupedaddress,amssymb,amsmath]{revtex4}

\usepackage{graphicx}
\usepackage{hyperref}
\usepackage[all]{hypcap}
\hypersetup{
	colorlinks=true,        %
    linkcolor=blue,         %
    citecolor=red,        %
    filecolor=magenta,      %
    urlcolor=blue
}

\def\Srsix{$^{86}$Sr$^+$}

\def\Srseven{$^{87}$Sr$^+$}

\def\Sreight{$^{88}$Sr$^+$}

\newcommand{\ket}[1]{|#1\rangle}

\begin{document}

\title{Precision measurement of the $5^{2}S_{1/2}$ - $4^{2}D_{5/2}$ quadrupole transition isotope shift between \Sreight \ and \Srsix}

\author{Warren E. Lybarger, Jr.}
\email[e-mail:\ ]{weljr@lanl.gov}
\affiliation{Los Alamos National Laboratory, Applied Modern Physics Group, MS D434, Los Alamos, NM 87545, USA}
\affiliation{Department of Physics and Astronomy, University of California, Los Angeles, CA 90095, USA}

\author{Julian C. Berengut}
\email[e-mail:\ ]{jcb@phys.unsw.edu.au}
\affiliation{School of Physics, University of New South Wales, Sydney 2052, Australia}

\author{John Chiaverini}
\email[e-mail:\ ]{john.chiaverini@ll.mit.edu}
\altaffiliation{current address: MIT Lincoln Laboratory}
\affiliation{Los Alamos National Laboratory, Applied Modern Physics Group, MS D434, Los Alamos, NM 87545, USA}

\date{\today}

\begin{abstract}

We have measured the isotope shift of the narrow quadrupole-allowed $5^{2}S_{1/2}$ - $4^{2}D_{5/2}$ transition in \Srsix \ relative to the most abundant isotope \Sreight.  This was accomplished using high-resolution laser spectroscopy of individual trapped ions, and the measured shift is $\Delta\nu_\textrm{meas}^{88,86}=570.281(4)$~MHz.  We have also tested a recently developed and successful method for \emph{ab-initio} calculation of isotope shifts in alkali-like atomic systems against this measurement, and our initial result of $\Delta \nu_\textrm{calc}^{88,86} = 457\,(28)$~MHz is also presented.  To our knowledge, this is the first high precision measurement and calculation of that isotope shift.  While the measurement and the calculation are in broad agreement, there is a clear discrepancy between them, and we believe that the specific mass shift was underestimated in our calculation.  Our measurement provides a stringent test for further refinements of theoretical isotope shift calculation methods for atomic systems with a single valence electron.

\end{abstract}

\maketitle

\section{Introduction\label{sec_intro}}
Isotope shift measurements provide precise tests of our understanding of atomic and nuclear structure \cite{PhysRevA.67.013402, PhysRevC.32.2058, PhysRevLett.60.2607, wang2007, zhao95a, PhysRevA.60.2867}.  %
Additionally, precise knowledge of isotope shifts is useful in studying alternative explanations for observations of an apparent temporal variation of the fine structure constant~\cite{1berengut03pra}.  The problem of isotope shift calculation is rich in relativistic many-body physics, and it has historically presented a strong challenge when considering atomic species with many electrons.  Motivated by the aforementioned applications, there has been much recent work toward improving theoretical methods for calculating isotope shifts~\cite{1berengut03pra, 2berengut06pra, 3berengut08jpb}.  Along with others~\cite{zhao95a,barwood97a,PhysRevA.67.013402}, the measurement reported here provides a stringent test case for our calculation method~\cite{1berengut03pra} and other suitable methods in the context of ions with alkali-like electronic structure.  In this paper, we present a high precision measurement of the $A=88$ to $A=86$ isotope shift in the $S_{1/2}$ - $D_{5/2}$ quadrupole transition of Sr II, a medium-weight ion, and we also test the \emph{ab-initio} calculation method described in~\cite{1berengut03pra} with this measurement.  %
The isotope shift of this quadrupole transition in Sr II has proven to be in a particularly interesting regime for theoretical work, and, to our knowledge, the calculation presented here is a first reported calculation of this shift.

The experimental technique used here involves quantum-shelving spectroscopy of individual ions, in which kilohertz-level determinations are possible due to the trapping of an atom in a potential with a strength significantly larger than the recoil energy of the photons used to interrogate it~\cite{zhao95a}.  The very high spectroscopic precision of the measurement is also afforded, in part, by the use of a dipole-forbidden transition with a linewidth that is approximately eight orders of magnitude smaller than the more commonly studied dipole-allowed transitions in Sr II~\cite{heilig1961, hughes1957, Lorenzen198226, PhysRevC.32.2058, PhysRevLett.60.2607}.  With the natural linewidth of this transition being a mere 0.46~Hz~\cite{gerz87, james98a}, the precision of the present measurement is limited by the frequency stability of the probe laser used to interogate the quadrupole transition.  To our knowledge, the most precise experimental determination of the \Sreight \ - \Srsix \ isotope shift in the $S_{1/2}$ - $D_{5/2}$ transition prior to the present one was made at the megahertz level~\cite{barwood97a} (though from the format in which the result is presented, this prior determination may only be at the 10~MHz level; no uncertainty is quoted).  Our measurement improves upon this precision by more than two orders of magnitude.  We also note that a similar measurement of 247.99(4)~MHz for the \Sreight \ - \Srseven \ isotope shift of the same quadrupole transition has also been reported in the context of a \Srseven \ ion optical frequency standard~\cite{PhysRevA.67.013402}.

\begin{figure}
\includegraphics[width=1 \columnwidth]{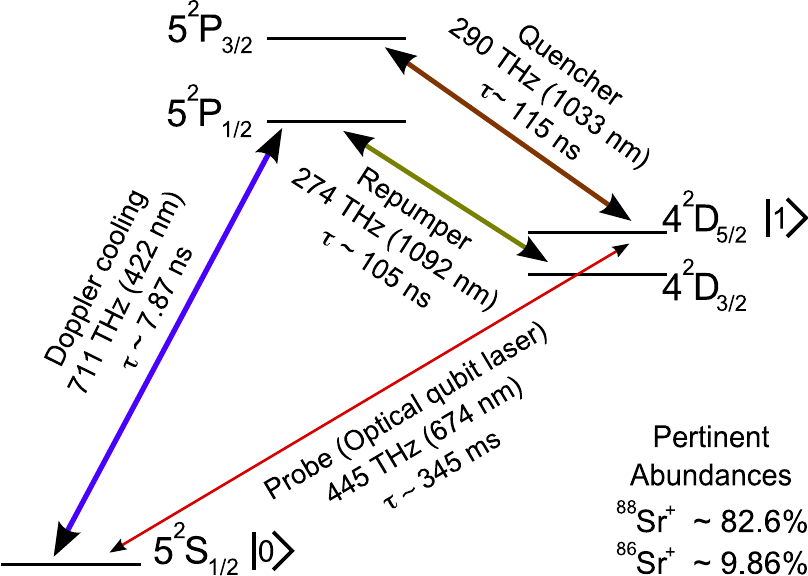}
\caption{(Color online) Level diagram for even isotopes of singly ionized strontium (i.e., Sr II) with no net nuclear spin, \Sreight\ and \Srsix \ in this context.  The $S_{1/2}$ level is the ground state of the single valence electron in these ions.  The $P$ levels decay rapidly, having lifetimes of approximately ten nanoseconds, and the $D$ levels are metastable (lifetime approximately half a second).  The double-arrows indicate laser radiation used in the determination of the isotope shift in the 445~THz (red) quadrupole transition. The thick lines represent dipole-allowed ($E1$) transitions while the thin line indicates the much weaker dipole-forbidden but quadrupole-allowed ($E2$) transition of interest. The natural lifetimes $\tau$ of the excited states are taken from \cite{james98a}. The eigenstates of the optical qubit in the quantum information processing apparatus used to perform this measurement are $\ket{0}=\ket{S_{1/2}}$\ and $\ket{1}=\ket{D_{5/2}}$.\label{fig_levels}}
\end{figure}

Aside from providing a very stringent test for atomic and nuclear structure calculations in atomic systems possessing alkali-like electronic structure, our measurement also provides a useful tag for identification of \Srsix\ in ion frequency standards or quantum information processing applications, especially when combined with absolute frequency measurements~\cite{PhysRevA.67.032501, Margolis04} for the same transition in the more abundant and commonly used isotope \Sreight.  One application in which this tag might be useful is in investigations of sympathetic sideband cooling of one or more motional modes of a multi-isotope coulomb crystal, where one ion species is utilized for ``memory qubits'' and the other is used as ``refrigerator'' ions~\cite{PhysRevA.61.032310,rohde01a,blinov02a,barrett03a,PhysRevA.79.050305,lybarger10}.  This may be of intereset as the isotope shift reported here may be large enough for use of laser radiation derived from the same 445~THz source to simultaneously perform both sympathetic near-ground-state cooling and coherent operations on memory qubits in a manner similar to the work of~\cite{PhysRevA.79.050305}, i.e., with an optical qubit encoded in the $S_{1/2}$ and $D_{5/2}$ levels (cf. Fig. \ref{fig_levels}).

A detailed account of our experimental methods is presented in section \ref{sec_methods} followed by the results in section \ref{sec_results}.  The discussion proceeds with pertinent details of the calculation of the quadrupole transition isotope shift in section \ref{sec_calcs}.  This report concludes with a comparison of and remarks on our theoretical and experimental results in section \ref{sec_conclusions}.

\section{Experimental Methods\label{sec_methods}}

We trap Sr$^+$ ions in a four-rod radio frequency (RF) Paul trap, described previously~\cite{berkeland02a, lybarger10}, with axial and radial trapping frequencies of $\sim\nobreak500$~kHz and $\sim\nobreak1.2$~MHz, respectively.  Ions are loaded one-by-one from a neutral strontium atomic vapor via photoionization (PI) using a two step process to an autoionizing level with laser radiation at 650~THz and 740~THz~\cite{vant06a}.  Ions are Doppler cooled to approximately the Doppler limit using light of frequency 711~THz, slightly red-detuned from the $S_{1/2}$ - $P_{1/2}$ dipole-allowed cycling transition.  We have noticed a small amount of isotope selectivity during ion loading by tuning the first-step PI laser slightly to the red by a few hundred megahertz so that we can load $^{86}$Sr$^+$ from our unenriched strontium sample somewhat more often than would be expected from the natural abundances of strontium isotopes ($9.86\%$ $^{86}$Sr$^+$; $82.6\%$ $^{88}$Sr$^+$).  %
We occasionally load $^{87}$Sr$^+$ as well.  However, our current apparatus does not allow direct cooling or detection of this species due to its ground-state hyperfine splitting.  The $P_{1/2}$ level decays to the $D_{3/2}$ level, which has a 0.4~s lifetime, approximately 1 in 14 times.  Consequently, a ``repumping'' laser at 274~THz continuously illuminates the ion to depopulate this level and return the valence electron to the cooling cycle (cf. Fig. ~\ref{fig_levels}).  The ion is in the Lamb-Dicke regime~\cite{PhysRev.89.472} for the quadrupole-transition probe light at 445~THz, thus allowing high resolution first-order Doppler-shift-free spectroscopy on this narrow transition via quantum shelving spectroscopy of the ion~\cite{PhysRevLett.56.2797}.

The 445~THz light is generated using a homebuilt external-cavity diode laser that is narrowed and stabilized with a Zerodur low-thermal-expansion optical reference cavity.  The cavity is housed in a vacuum chamber in a thermally controlled enclosure with temperature stability at the millikelvin level, and the laser is locked to this cavity using the Pound-Drever-Hall technique~\cite{drever83a}.  Feedback is provided both to the external laser cavity grating and to the laser diode current to narrow and frequency lock the laser. This has allowed a demonstrated lower bound on the optical qubit (cf. Fig. \ref{fig_levels}) coherence time of $2.5(0.2)$ ms when using simple refocusing techniques to circumvent technical and environmental noise (e.g., magnetic field fluctuations at the ion location)~\cite{lybarger10}.  On the much longer timescale of the several seconds that are required to perform precision spectroscopy on this optical qubit, the probe laser frequency varies due to drifts in both the length of its reference cavity and the locking electronics.  On this timescale, the demonstrated linewidth is still less than 1~kHz, thus providing for the observation of narrow quadrupole transition spectral lines~\cite{lybarger10}.  After each interrogation on this transition and subsequent state determination, the ion state is reset to the ground state $S_{1/2}$ level by driving any population in the $D_{5/2}$ level to the $P_{3/2}$ level from where it predominantly decays to the ground state.  This ``quenching'' is accomplished using a 290~THz laser (cf. Fig. ~\ref{fig_levels}).  By applying this radiation simultaneously with the 274~THz repumping light, the ion will be optically pumped to the $S_{1/2}$ state with very high probability in several tens of nanoseconds.  The 445~THz spectroscopy beam is pulsed using an acousto-optic modulator (AOM) in a double-pass configuration, which allows for scanning of its frequency over a few tens of megahertz.  The cooling and quenching laser beams are switched off using shutters in order to avoid coupling of the $D_{5/2}$ level to other levels, which would otherwise broaden and shift the narrow quadrupole transition while it is being probed.

Each quantum shelving spectroscopy trial proceeds as follows.  The ion is first Doppler cooled for approximately 10~ms with the Doppler, repumping, and quenching laser radiation on.  The Doppler radiation and the quencher are then shuttered off.  Approximately 1~ms later, the narrow transition is probed via a pulse from the spectroscopy laser of varying frequency detuning and typically of 1 to 10~ms in duration.  Another few to several milliseconds later, the Doppler laser is switched back on, and after less than 1~ms, fluorescence is detected using a photomultiplier tube for 5~ms with near unity state detection efficiency.  If the ion was not excited by the spectroscopy pulse, it will fluoresce and a mean of approximately 35 Poisson-distributed photon counts are acquired in this detection time.  If the spectroscopy pulse succeeded in exciting the ion to the $D_{5/2}$ level where it then remains shelved (i.e., removed from the Doppler-cooling cycle) during detection, a mean of approximately 1.8 photons is collected.  This background is due primarily to 711~THz Doppler cooling light scattering from the trap electrodes into the fluorescence collection optics.  Finally, the quencher beam is switched on to pump the ion back to the ground state in case it was shelved, and the ion is then re-cooled for a few milliseconds prior to commencement of the following trial.  The repumper beam remains on during the entire experiment; the AC Stark shift and broadening due to this radiation on the narrow transition of interest are negligible because of its $\sim\nobreak16$~THz detuning from the closest transition involving the $D_{5/2}$ level.

\begin{figure}[tbp]
\includegraphics[width=1 \columnwidth]{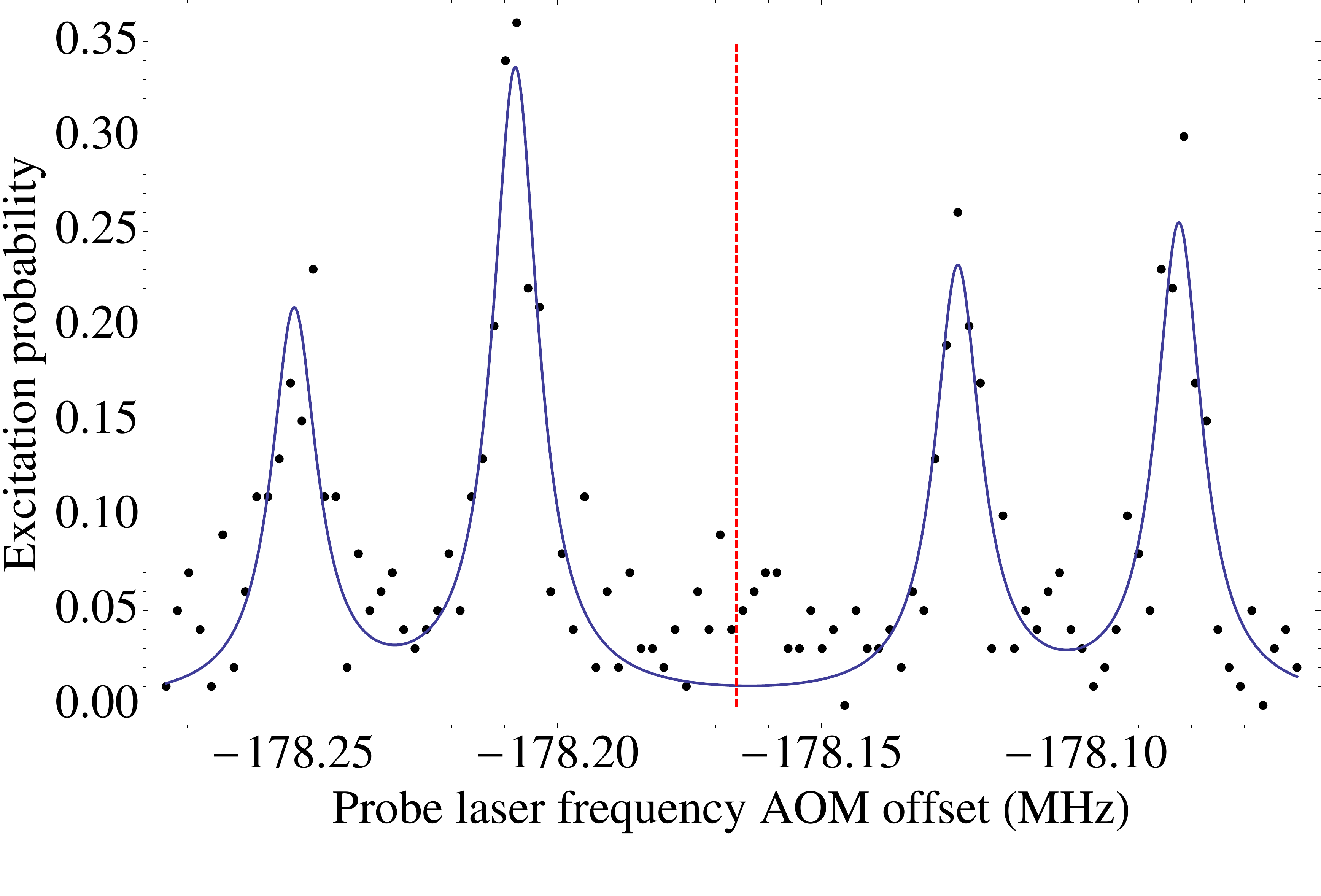}
\caption{(Color online) Sample spectrum of the $S_{1/2}$ - $D_{5/2}$ transition in Sr II acquired using the quantum shelving technique.  The blue curve is a fit to eight Zeeman components of this quadrupole transition with the four most central Zeeman lines shown here.  The dashed red line indicates the measured RF frequency delivered to the AOM at the transition center, which was determined as a parameter of the fit.\label{fig_spectrum}}
\end{figure}

At each frequency detuning of the probe laser, 100 trials were carried out.  A photon count threshold, typically of 9 counts, was set in order to minimize error in distinguishing dark signals from bright ones due to the overlap of their respective Poisson distributions \cite{myerson08a}, where trials with a photon count at or below the threshold were considered as dark and all others were bright.   
The ratio of trials where the ion was dark to the total number of trials at a given detuning gave the excitation probability at each frequency step.  Spectra like that shown in Fig.~\ref{fig_spectrum} were constructed in this manner, and they display the Zeeman components of the $S_{1/2}$ - $D_{5/2}$ transition.  
The laser detuning that corresponded to the center frequency of a given quadrupole transition spectrum was determined as a parameter of a fit to the spectral data.  The total magnetic field at the location of the ion was also determined from the Zeeman spectra, and each trial was triggered on the AC line to minimize effects from magnetic field fluctuations at the line frequency of 60~Hz~\footnote{Magnetic field fluctuations on timescales much longer and much shorter than the frequency scan duration do not significantly affect the central frequency determination; the former simply lead to a scaling of the line spacing, the latter to broadening of the lines.  Fluctuations on the few-minute timescale of a spectral scan can lead to asymmetry in the Zeeman structure, but our multiple-line fit accounts for this effect.  Furthermore, our measurements are consistent with no significant field fluctuations above the $10^{-7}$~T level on this timescale, and any related errors, on the order of 1 kHz or less, are subsumed into our fit errors.}.  We carried out these determinations for both isotopes of interest, and comparison of the RF frequencies delivered to the AOM at the transition centers enabled the determination of the isotope shift.

\begin{figure}[tbp]
\includegraphics[width=1 \columnwidth]{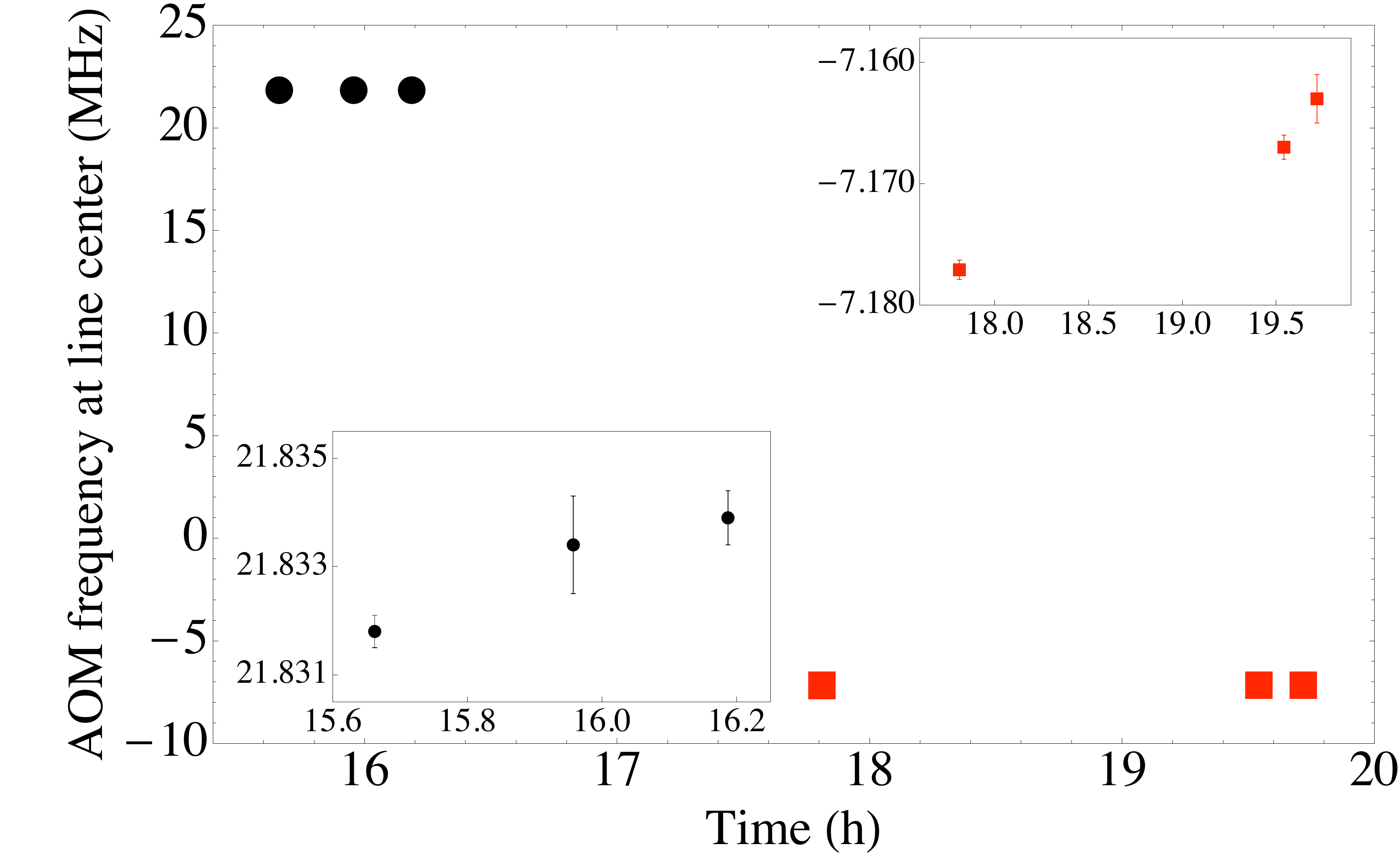}
\caption{(Color online) Measures of the AOM component $\Delta\nu^{88,86}_{AOM}$ of the overall isotope shift.  Black circles are measurements of the $S-D$ transition in \Srsix, and red squares are measurements of the same transition in \Sreight.  The AOM frequency is the RF frequency shift delivered to the probe light at the resonance line center not including the offset due to the free spectral range of the reference cavity that was used to lock and narrow the probe laser.  The laser was locked one cavity FSR to the red when measuring the \Srsix \ transition compared to when measurements of the \Sreight \ transition were performed.  %
The insets show the same points with zoomed ranges to clearly show the effect of cavity drift (note the different ordinate scales).  An overall frequency offset of 200 MHz has been subtracted from all RF frequency values for readability.\label{fig_bareshift}}
\end{figure}

The total shift is on the order of the approximately 600~MHz free spectral range (FSR) of the 445~THz probe laser reference cavity, and the two different cavity fringes used to lock that laser for probing the two isotopes were in adjacent FSR's as a result.  The full determination of the isotope shift therefore required a measurement of the FSR of the probe laser reference cavity as well.  The total shift in frequency detuning is the AOM component of the shift $\Delta\nu^{88,86}_{AOM}$ measured from the shelving method just described with subsequent subtraction of one FSR, $\Delta\nu_{FSR}$, of the reference cavity (i.e., the lock point for \Srsix\ is approximately one cavity FSR red of that for \Sreight):
\begin{equation}
\Delta\nu_\textrm{meas}^{88,86}= \Delta\nu^{88,86}_{AOM}-\Delta\nu_{FSR}.
\label{eq_totalshift}
\end{equation}
We note that the convention $\Delta\nu^{A',A}=\nu^{A'}-\nu^{A}$ is followed here, where $\nu^{A'}$ and $\nu^{A}$ are the energies of the transition of interest in the two isotopes with mass numbers $A'$ and $A$ with $A' > A$.

To determine $\Delta\nu_{FSR}$ with some degree of precision, we used an electro-optic modulator (EOM) in an auxiliary measurement to impose frequency sidebands at approximately half of the cavity FSR on the portion of the probe laser light that is coupled into the fixed-length reference cavity.  While scanning the grating of the laser, we monitored the transmission through the cavity; by adjusting the EOM drive frequency such that the first-order blue sideband from one FSR overlapped the first-order red sideband from an adjacent FSR, we could determine $\Delta\nu_{FSR}$ to a precision limited by the  laser linewidth and the cavity finesse at 445~THz.  We measured the reference cavity FSR to be 
\begin{equation}
\Delta\nu_{FSR}=599.301(2)~\textrm{MHz}.  
\label{eq_cavityFSR}
\end{equation}

The isotope shift, as measured using the shelving method, is affected by an isothermal drift in the reference cavity  length.  Through measurement of the transition resonance in individual ions over several months, we have determined the long-term average of the cavity drift to be 52(1)~kHz/d.  On shorter time scales, however, the cavity exhibits nonlinear drift, and it has been observed to drift at rates up to approximately six times the long-term drift rate over a few hours.  To date, our isotope-shift measurements have been performed on only one ion at a time.  This leads to the requirement that $\Delta\nu^{88,86}_{AOM}$ must be determined from the difference of two measurements taken at different times, with the isotope shift determined as a parameter of a fit to the nonlinear cavity drift as explained below.  Lack of detailed knowledge about this nonlinear drift as a function of time dominates the imprecision of our current determination.  It should be noted that this drift corresponds to a maximum of approximately 3.6~Hz/s.  As the frequency scan durations were 2 to 4 minutes, we estimate the error from cavity drift during a scan to be a few hundred hertz at most, which is significantly below our uncertainty in center frequency determination at the few kilohertz level, and it has been ignored in this analysis.  The cavity drift also affects $\Delta\nu_{FSR}$ through the change in the length of the cavity, but the fractional length change is approximately $1.2\times 10^{-10}$ per day, leading to a nominal drift in the FSR of 3~mHz/h (up to $\sim\nobreak9$~mHz/h), which is well below our uncertainties during all measurements contributing to the present isotope shift determination.

The data used in the analysis consist of (i) three measurements of the $S_{1/2}$ - $D_{5/2}$ transition frequency in \Srsix, (ii) three measurements of the same transition frequency in \Sreight, and (iii) the measurement of the probe laser reference cavity FSR, as described above.  The six $S_{1/2}$ - $D_{5/2}$ measurements were obtained over a span of approximately four hours, a timescale over which the nonlinear reference cavity drift has a minimal effect on our uncertainty.  The RF frequencies delivered to the probe light via the AOM at the centers of the acquired Zeeman spectra are plotted in Fig.~\ref{fig_bareshift} versus the relative time of each measurement (with an arbitrary zero of time at 00:00 local time on the day that the data were collected).  The zoomed insets show the effect of cavity drift between measurements.

To best account for the cavity drift when determining the AOM shift, $\Delta\nu^{88,86}_{AOM}$ was introduced as a fit parameter, and a simple drift function with a discontinuity was fit to the six measurement points versus time.  As the exact character of the cavity drift during data acquisition was not known, two different drift functions, one linear and one quadratic, were chosen:

\begin{widetext}
\begin{eqnarray}
f_1(t)&=& (m_1 t + b_1)\Theta(t_0-t)+(m_1 t + b_1 + \Delta\nu^{88,86}_{AOM})\Theta(t-t_0)\\
f_2(t)&=&(q_2 t^2 + m_2 t + b_2)\Theta(t_0-t)+(q_2 t^2 + m_2 t + b_2 + \Delta\nu^{88,86}_{AOM})\Theta(t-t_0).
\end{eqnarray}
\end{widetext}

\noindent Here $\Theta$ is the Heaviside unit step function, and the $q_i$, $m_i$, and $b_i$ are fit parameters for function $f_i$ of time $t$.  The discontinuity time $t_0$ can be any time between the \Srsix\ and \Sreight\ measurements, and it is chosen here to be 17~h.  The reference cavity drift is sufficiently slow such that a higher than quadratic order fit function is not required~\footnote{It is possible that the probe laser reference cavity could have undergone a non-deterministic (e.g. not linear or quadratic) drift rate for some period during the approximately 90 minute interval between the last \Srsix \ measurement and the first \Sreight \ measurement. Such a drift would have led to a systematic error in our estimate of $\Delta\nu^{88,86}_{AOM}$ by the above analysis. We cannot eliminate this possibility with certainty since a \textit{second} absolute frequency reference for continuously monitoring the cavity drift was unavailable when the data presented in Fig. 3 were acquired. However, an extensive auxiliary analysis of $S-D$ spectra taken during the period of a few months preceding and following our primary measurement indicates that on the relevant timescales (i) the likelihood of such an event having occurred is quite remote and (ii) the maximum systematic error that would have accrued in the isotope shift is approximately the size of our quoted uncertainty. For these reasons, we have not included this effect separately in our error estimate.}.

\section{Discussion of Experimental Results\label{sec_results}}

\begin{figure}[tbp]
\includegraphics[width=1 \columnwidth]{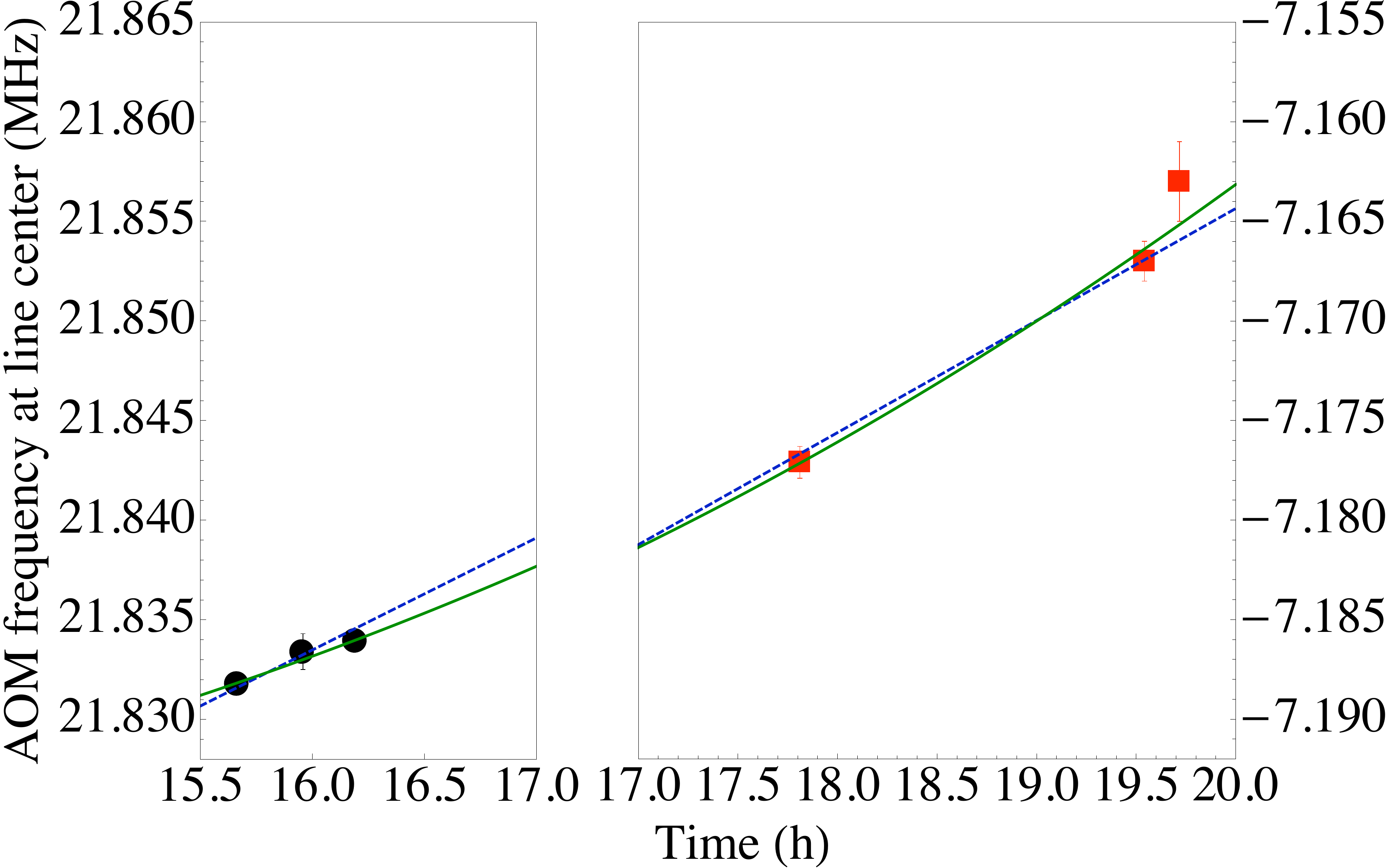}
\caption{(Color online) Fits to determine AOM component of isotope shift.  Black circles are measurements of the $S-D$ transition in \Srsix, and red squares are measurements of the same transition in \Sreight.  The blue dashed line is a linear fit yielding an AOM shift of 29.020(2)~MHz, and the green solid line is a quadratic fit, yielding an AOM shift of 29.019(2)~MHz.  The two panels are offset in the ordinate coordinate by 29.020~MHz, but are scaled identically (refer to left and right scales), as are the abscissas.  An overall frequency offset of 200 MHz has been subtracted from all RF frequency values for readability.\label{fig_shiftfit}}
\end{figure}

The linear fit to $f_1$ yields $\Delta\nu^{88,86}_{AOM}=29.0204(21)$~MHz with a reduced $\chi^2$ (goodness-of-fit parameter) of 1.54 (3 degrees of freedom).  The quadratic fit to $f_2$ yields $\Delta\nu^{88,86}_{AOM}=29.0191(18)$~MHz with a reduced $\chi^2$ of 0.88 (2 degrees of freedom); see Fig.~\ref{fig_shiftfit}.  As the actual drift function during these measurements is unknown, we weight the average of these two measurements of $\Delta\nu^{88,86}_{AOM}$ by the inverses of their respective goodness-of-fit parameters.  This gives a value of $\Delta\nu^{88,86}_{AOM}=29.020(3)$~MHz, where the uncertainty has been increased from 0.002~MHz to 0.003~MHz to conservatively account for lack of knowledge of the drift function.  With the value for $\Delta\nu_{FSR}$ in equation (\ref{eq_cavityFSR}), we calculate a value (cf. equation \ref{eq_totalshift}) for the quadrupole transition isotope shift to be

\begin{equation}
\Delta\nu_\textrm{meas}^{88,86}=570.281(4) {\rm MHz.}
\label{eq_totalShift}
\end{equation} 

\noindent To our knowledge, this is the first measurement of this shift to high precision, at a level of better than 1 part in 10$^5$, with the uncertainty dominated by nonlinear drift in the probe laser reference cavity.  From equation (\ref{eq_totalShift}) and the known value of the absolute frequency of the $5^{2}S_{1/2}$ - $4^{2}D_{5/2}$ transition in \Sreight~\cite{Margolis04}, we also infer an absolute frequency for this transition in \Srsix \ of
\begin{equation}
\nu_{86} = 444,778,473,814(4) {\rm kHz}.
\label{eq_absFreqSrsix}
\end{equation}

\section{Calculation of the Isotope Shift\label{sec_calcs}}

To an excellent approximation, isotope shifts in atomic transition frequencies arise from two sources~\cite{woodgate1983, king1984, heilig1974}: (i) the mass shift (MS) and (ii) the field shift (FS).  The MS is the difference in recoil energy between isotopes due to their differing nuclear masses; i.e., when conserving momentum in the atomic systems, the eigenenergies of their electronic states shift to account for the differing recoil of the nuclei.  The MS is itself comprised of the sum of two different effects referred to as the ``normal mass shift'' (NMS) and the ``specific mass shift'' (SMS).  The NMS is due to a reduced mass correction which accounts for the finite mass of a real atomic nucleus compared to the infinitely massive nucleus of an idealized atom, and it is readily calculated.  The SMS accounts for the modification of the nuclear recoil energy due to correlations in the motions of all possible pairs of electrons orbiting the nucleus via conservation of momentum applied to the atomic system.  Accurate calculation of this shift is a difficult problem that depends heavily on the accuracy with which the multi-electron wavefunction for the atom can be determined.  The FS is a consequence of the difference in the neutron-number-dependent size and shape of the nuclear electric charge distribution between isotopes; i.e., the eigenenergies of the electronic states adjust to the unique electric potentials of their nuclei.  %
Measurement of the FS requires the use of an electronic transition that involves a significant change in the electronic wavefunction density that penetrates the nuclear volume (or in approximate terms, $\Delta\left|\psi(r=0)\right|^2 \neq 0$, where $r=0$ is the nuclear center and $\psi$ is the electronic wavefunction).  The electronic states most sensitive to nuclear screening are the $S$ states followed by the $P$ states, with states in all other angular momentum levels having negligible penetration of the nuclear volume~\cite{woodgate1983}.  Thus, in principle, the quadrupole transition utilized here should be a sensitive probe for measuring the FS in this medium-weight atomic ion where neither the SMS or FS may be treated as negligible \textit{a priori}.

The difference in the transition frequency $\Delta\nu^{A',A}$ between isotopes with mass numbers $A'$ and $A$ can be expressed as~\cite{1berengut03pra}
\begin{equation}
\Delta \nu^{A',A} = (k_\mathit{NMS} + k_\mathit{SMS}) \left(\frac{1}{A'} - \frac{1}{A}\right)
	+ F \delta\!\left< r^2 \right>^{A', A} .
\label{eq_isEquation}
\end{equation}
Here, the normal mass shift constant is
\[
	k_\mathit{NMS} = -\frac{\nu}{1823} ,
\]
where $\nu$ is the transition frequency of the idealized atom, 1823 refers to the ratio of the atomic mass unit to the electron mass, and $\delta\!\left< r^2 \right>^{A', A}$ is the relative difference in the mean squared nuclear charge radius between the two nuclear species. The SMS and FS constants, $k_\mathit{SMS}$ and $F$, must be calculated.

The Sr~II ion considered in this paper can be treated as a single valence electron above a closed-shell electronic core. Therefore the methods of ref.~\cite{1berengut03pra} for computing $F$ and $k_\mathit{SMS}$ are appropriate. Full details are given in \cite{1berengut03pra}; here we only present an outline. The general idea is to include an additional term which represents the SMS or FS --- a ``finite field'' --- in an energy calculation from the very beginning. We rescale the finite field using a parameter $\lambda$, chosen to make the effect large but numerically stable. The energy calculation is then repeated for different values of $\lambda$ and the numerical constant is extracted as the gradient, i.e.~$k_\mathit{SMS} = dE/d\lambda$.

Calculation of the energy is done with the AMBiT code~\cite{2berengut06pra,3berengut08jpb}. We use the Dirac-Fock method to generate the core orbitals and their potential. Valence and virtual orbitals are then constructed via a large $B$-spline basis~\cite{4johnson88pra} so that core-valence correlations can be included using many-body perturbation theory (MBPT). We calculate these correlations to second order in the operator $H - H_{DF}$, the difference between the exact and Dirac-Fock Hamiltonians. Higher-order correlations are estimated using Brueckner orbitals. An operator $\hat \Sigma$ is created such that the second-order MBPT correction to a valence orbital $\left|n\right>$ is $\delta E^{(2)}_n = \left< n| \hat \Sigma |n\right>$. Then $\hat \Sigma$ is added to the exchange potential in the Dirac-Fock Hamiltonian for the valence electron and this new potential is used to create the single-electron Brueckner orbital. We use the difference between the second-order solution and the Brueckner orbital solution as an estimate of the errors associated with higher-order correlations.

For the $S_{1/2}$ - $D_{5/2}$ transition in Sr~II we obtain
\begin{align*}
k_\mathit{NMS} &= -244\textrm{~GHz}\cdot\textrm{amu,}\\
k_\mathit{SMS} &= -1149\,(50)\textrm{~GHz}\cdot\textrm{amu, and}\\
F &= -1616\,(272)\textrm{~MHz/fm}^2.
\end{align*}
From this intermediate result, we calculate a total mass shift of $368\,(13)$~MHz and a field shift of $89\,(15)$~MHz, where we use $\delta\!\left< r^2 \right>^{88,86} = -0.0549\,(17)$~fm$^2$ from ~\cite{5angeli04adndt}.  Adding these two contributions (cf. equation \ref{eq_isEquation}), we finally obtain
\begin{equation}
\Delta \nu_\textrm{calc}^{88,86} = 457\,(28)~\textrm{MHz}.
\end{equation}

Since we have measurements for $\Delta\nu_\textrm{meas}^{88,86}$ (this work) and $\Delta\nu_\textrm{meas}^{88,87}$~\cite{PhysRevA.67.013402} for the same quadrupole transition in Sr II, we may also use equation (\ref{eq_isEquation}) and the appropriate $\delta\!\left< r^2 \right>^{88,x}$ from~\cite{5angeli04adndt} to determine $k_\mathit{SMS}$ and $F$ experimentally.  These constants are determined to be 
\begin{align*}
k_\mathit{SMS} &= -1660\,(20)\textrm{~GHz}\cdot\textrm{amu and}\\
F &= -1200\,(100)\textrm{~MHz/fm}^2.
\end{align*}
We see that our ab-initio calculation has significantly underestimated the magnitude of the specific mass shift in this ion, a discrepancy which is not accounted for by the error estimation method.

\section{Summary and Concluding Remarks\label{sec_conclusions}}

We have presented a high precision measurement, to better than 1 part in 10$^5$, of the \Sreight \ - \Srsix \ isotope shift of the $5^{2}S_{1/2}$ - $4^{2}D_{5/2}$ quadrupole transition of Sr II that is consistent with the result reported in~\cite{barwood97a} and which improves upon that reported precision by more than two orders of magnitude.  %
The recently developed and successful method~\cite{1berengut03pra} for the \emph{ab-initio} calculation of isotope shifts in single electron atomic systems has also been tested against our new measurement, and we use the present measurement in conjunction with an analogous measurement of the \Sreight - \Srseven \ isotope shift~\cite{PhysRevA.67.013402} to perform an experimental check on our calculation.  Sr II is a medium-weight atomic ion that presents a somewhat difficult case for theoretical calculations of its isotope shifts since the SMS and FS are both large, and neither one may be regarded as negligible, thus making it difficult to separate these two quantities in equation (\ref{eq_isEquation}) without some independent information about either one of them.  There is a clear discrepancy between our calculated isotope shift and the measured values. The error is most likely to be in the SMS, since the operator for this effect, \mbox{$\Sigma_{i < j} \mathbf{p}_i \cdot \mathbf{p}_j$}, $\mathbf{p}_k$ being the momentum of electron $k$, is very sensitive to details of the electronic wavefunctions. Relativistic corrections to the isotope shift are at the level of a few percent in light atoms~\cite{6korol07pra}, and should not be much larger in our case.  Therefore, the most likely reason for the discrepancy is higher-order core-valence correlations (beyond second order) that are not included in our calculation. Interestingly, our method for estimating the magnitude of higher-order correlations (using Brueckner orbitals) has failed in this case: the calculated $k_\mathit{SMS}$ is 31\% too small, but our predicted error was 6\%. Nevertheless, the calculation is in broad agreement with the measurement, which is a promising initial result when one considers that the specific mass shift is unusually dominant in this case; it is approximately 7 times the size of the normal mass shift.  Our high precision measurement presents a strong test for further refinements in theoretical iosotope shift calculation techniques for alkali-like atomic systems.

\begin{acknowledgments}

We gratefully acknowledge Dana Berkeland's efforts in building the ion trap and the interrogation laser used in these determinations and Bob Scarlett for help in improving the long-term stability of that laser.  We thank Malcolm Boshier for helpful comments during the manuscript preparation and Xinxin Zhao for the use of the electro-optic modulator for the measurement of the free spectral range of the interogation laser reference cavity.  This work has been supported by DOE through the LANL Laboratory Directed Research and Development program.

\end{acknowledgments}

\bibliography{ionbasic2}

\begin{thebibliography}{35}
\expandafter\ifx\csname natexlab\endcsname\relax\def\natexlab#1{#1}\fi
\expandafter\ifx\csname bibnamefont\endcsname\relax
  \def\bibnamefont#1{#1}\fi
\expandafter\ifx\csname bibfnamefont\endcsname\relax
  \def\bibfnamefont#1{#1}\fi
\expandafter\ifx\csname citenamefont\endcsname\relax
  \def\citenamefont#1{#1}\fi
\expandafter\ifx\csname url\endcsname\relax
  \def\url#1{\texttt{#1}}\fi
\expandafter\ifx\csname urlprefix\endcsname\relax\def\urlprefix{URL }\fi
\providecommand{\bibinfo}[2]{#2}
\providecommand{\eprint}[2][]{\url{#2}}

\bibitem[{\citenamefont{Barwood et~al.}(2003)\citenamefont{Barwood, Gao, Gill,
  Huang, and Klein}}]{PhysRevA.67.013402}
\bibinfo{author}{\bibfnamefont{G.~P.} \bibnamefont{Barwood}},
  \bibinfo{author}{\bibfnamefont{K.}~\bibnamefont{Gao}},
  \bibinfo{author}{\bibfnamefont{P.}~\bibnamefont{Gill}},
  \bibinfo{author}{\bibfnamefont{G.}~\bibnamefont{Huang}}, \bibnamefont{and}
  \bibinfo{author}{\bibfnamefont{H.~A.} \bibnamefont{Klein}},
  \bibinfo{journal}{Phys. Rev. A} \textbf{\bibinfo{volume}{67}},
  \bibinfo{pages}{013402} (\bibinfo{year}{2003}).

\bibitem[{\citenamefont{Buchinger et~al.}(1985)\citenamefont{Buchinger,
  Corriveau, Ramsay, Berdichevsky, and Sprung}}]{PhysRevC.32.2058}
\bibinfo{author}{\bibfnamefont{F.}~\bibnamefont{Buchinger}},
  \bibinfo{author}{\bibfnamefont{R.}~\bibnamefont{Corriveau}},
  \bibinfo{author}{\bibfnamefont{E.~B.} \bibnamefont{Ramsay}},
  \bibinfo{author}{\bibfnamefont{D.}~\bibnamefont{Berdichevsky}},
  \bibnamefont{and} \bibinfo{author}{\bibfnamefont{D.~W.~L.}
  \bibnamefont{Sprung}}, \bibinfo{journal}{Phys. Rev. C}
  \textbf{\bibinfo{volume}{32}}, \bibinfo{pages}{2058} (\bibinfo{year}{1985}).

\bibitem[{\citenamefont{Silverans et~al.}(1988)\citenamefont{Silverans,
  Lievens, Vermeeren, Arnold, Neu, Neugart, Wendt, Buchinger, Ramsay, and
  Ulm}}]{PhysRevLett.60.2607}
\bibinfo{author}{\bibfnamefont{R.~E.} \bibnamefont{Silverans}},
  \bibinfo{author}{\bibfnamefont{P.}~\bibnamefont{Lievens}},
  \bibinfo{author}{\bibfnamefont{L.}~\bibnamefont{Vermeeren}},
  \bibinfo{author}{\bibfnamefont{E.}~\bibnamefont{Arnold}},
  \bibinfo{author}{\bibfnamefont{W.}~\bibnamefont{Neu}},
  \bibinfo{author}{\bibfnamefont{R.}~\bibnamefont{Neugart}},
  \bibinfo{author}{\bibfnamefont{K.}~\bibnamefont{Wendt}},
  \bibinfo{author}{\bibfnamefont{F.}~\bibnamefont{Buchinger}},
  \bibinfo{author}{\bibfnamefont{E.~B.} \bibnamefont{Ramsay}},
  \bibnamefont{and} \bibinfo{author}{\bibfnamefont{G.}~\bibnamefont{Ulm}},
  \bibinfo{journal}{Phys. Rev. Lett.} \textbf{\bibinfo{volume}{60}},
  \bibinfo{pages}{2607} (\bibinfo{year}{1988}).

\bibitem[{\citenamefont{Wang et~al.}(2007)\citenamefont{Wang, Dumke, Zhang,
  Liu, Stejskal, Zhao, Lu, Wang, Becker, and Walther}}]{wang2007}
\bibinfo{author}{\bibfnamefont{Y.~H.} \bibnamefont{Wang}},
  \bibinfo{author}{\bibfnamefont{R.}~\bibnamefont{Dumke}},
  \bibinfo{author}{\bibfnamefont{J.}~\bibnamefont{Zhang}},
  \bibinfo{author}{\bibfnamefont{T.}~\bibnamefont{Liu}},
  \bibinfo{author}{\bibfnamefont{A.}~\bibnamefont{Stejskal}},
  \bibinfo{author}{\bibfnamefont{Y.~N.} \bibnamefont{Zhao}},
  \bibinfo{author}{\bibfnamefont{Z.~H.} \bibnamefont{Lu}},
  \bibinfo{author}{\bibfnamefont{L.~J.} \bibnamefont{Wang}},
  \bibinfo{author}{\bibfnamefont{T.}~\bibnamefont{Becker}}, \bibnamefont{and}
  \bibinfo{author}{\bibfnamefont{H.}~\bibnamefont{Walther}},
  \bibinfo{journal}{European Physical Journal D} \textbf{\bibinfo{volume}{44}},
  \bibinfo{pages}{307} (\bibinfo{year}{2007}).

\bibitem[{\citenamefont{Zhao et~al.}(1995)\citenamefont{Zhao, Yu, Dehmelt, and
  Nagourney}}]{zhao95a}
\bibinfo{author}{\bibfnamefont{X.}~\bibnamefont{Zhao}},
  \bibinfo{author}{\bibfnamefont{N.}~\bibnamefont{Yu}},
  \bibinfo{author}{\bibfnamefont{H.}~\bibnamefont{Dehmelt}}, \bibnamefont{and}
  \bibinfo{author}{\bibfnamefont{W.}~\bibnamefont{Nagourney}},
  \bibinfo{journal}{Phys. Rev. A} \textbf{\bibinfo{volume}{51}},
  \bibinfo{pages}{4483} (\bibinfo{year}{1995}).

\bibitem[{\citenamefont{Roberts et~al.}(1999)\citenamefont{Roberts, Taylor,
  Gateva-Kostova, Clarke, Rowley, and Gill}}]{PhysRevA.60.2867}
\bibinfo{author}{\bibfnamefont{M.}~\bibnamefont{Roberts}},
  \bibinfo{author}{\bibfnamefont{P.}~\bibnamefont{Taylor}},
  \bibinfo{author}{\bibfnamefont{S.~V.} \bibnamefont{Gateva-Kostova}},
  \bibinfo{author}{\bibfnamefont{R.~B.~M.} \bibnamefont{Clarke}},
  \bibinfo{author}{\bibfnamefont{W.~R.~C.} \bibnamefont{Rowley}},
  \bibnamefont{and} \bibinfo{author}{\bibfnamefont{P.}~\bibnamefont{Gill}},
  \bibinfo{journal}{Phys. Rev. A} \textbf{\bibinfo{volume}{60}},
  \bibinfo{pages}{2867} (\bibinfo{year}{1999}).

\bibitem[{\citenamefont{Berengut et~al.}(2003)\citenamefont{Berengut, Dzuba,
  and Flambaum}}]{1berengut03pra}
\bibinfo{author}{\bibfnamefont{J.~C.} \bibnamefont{Berengut}},
  \bibinfo{author}{\bibfnamefont{V.~A.} \bibnamefont{Dzuba}}, \bibnamefont{and}
  \bibinfo{author}{\bibfnamefont{V.~V.} \bibnamefont{Flambaum}},
  \bibinfo{journal}{Phys. Rev. A} \textbf{\bibinfo{volume}{68}},
  \bibinfo{pages}{022502} (\bibinfo{year}{2003}).

\bibitem[{\citenamefont{Berengut et~al.}(2006)\citenamefont{Berengut, Flambaum,
  and Kozlov}}]{2berengut06pra}
\bibinfo{author}{\bibfnamefont{J.~C.} \bibnamefont{Berengut}},
  \bibinfo{author}{\bibfnamefont{V.~V.} \bibnamefont{Flambaum}},
  \bibnamefont{and} \bibinfo{author}{\bibfnamefont{M.~G.}
  \bibnamefont{Kozlov}}, \bibinfo{journal}{Phys. Rev. A}
  \textbf{\bibinfo{volume}{73}}, \bibinfo{pages}{012504}
  (\bibinfo{year}{2006}).

\bibitem[{\citenamefont{Berengut et~al.}(2008)\citenamefont{Berengut, Flambaum,
  and Kozlov}}]{3berengut08jpb}
\bibinfo{author}{\bibfnamefont{J.~C.} \bibnamefont{Berengut}},
  \bibinfo{author}{\bibfnamefont{V.~V.} \bibnamefont{Flambaum}},
  \bibnamefont{and} \bibinfo{author}{\bibfnamefont{M.~G.}
  \bibnamefont{Kozlov}}, \bibinfo{journal}{J. Phys. B}
  \textbf{\bibinfo{volume}{41}}, \bibinfo{pages}{235702}
  (\bibinfo{year}{2008}).

\bibitem[{\citenamefont{Barwood et~al.}(1997)\citenamefont{Barwood, Gill,
  Klein, and Rowley}}]{barwood97a}
\bibinfo{author}{\bibfnamefont{G.~P.} \bibnamefont{Barwood}},
  \bibinfo{author}{\bibfnamefont{P.}~\bibnamefont{Gill}},
  \bibinfo{author}{\bibfnamefont{H.~A.} \bibnamefont{Klein}}, \bibnamefont{and}
  \bibinfo{author}{\bibfnamefont{W.~R.~C.} \bibnamefont{Rowley}},
  \bibinfo{journal}{IEEE Trans. Instrum. Meas.} \textbf{\bibinfo{volume}{46}},
  \bibinfo{pages}{133} (\bibinfo{year}{1997}).

\bibitem[{\citenamefont{Heilig}(1961)}]{heilig1961}
\bibinfo{author}{\bibfnamefont{K.}~\bibnamefont{Heilig}}, \bibinfo{journal}{Z.
  Phys.} \textbf{\bibinfo{volume}{161}}, \bibinfo{pages}{252}
  (\bibinfo{year}{1961}).

\bibitem[{\citenamefont{Hughes}(1957)}]{hughes1957}
\bibinfo{author}{\bibfnamefont{R.~H.} \bibnamefont{Hughes}},
  \bibinfo{journal}{Phys. Rev.} \textbf{\bibinfo{volume}{105}},
  \bibinfo{pages}{1260} (\bibinfo{year}{1957}).

\bibitem[{\citenamefont{Lorenzen and Niemax}(1982)}]{Lorenzen198226}
\bibinfo{author}{\bibfnamefont{C.-J.} \bibnamefont{Lorenzen}} \bibnamefont{and}
  \bibinfo{author}{\bibfnamefont{K.}~\bibnamefont{Niemax}},
  \bibinfo{journal}{Optics Communications} \textbf{\bibinfo{volume}{43}},
  \bibinfo{pages}{26 } (\bibinfo{year}{1982}).

\bibitem[{\citenamefont{{Ch. Gerz} et~al.}(1987)\citenamefont{{Ch. Gerz}, {Th.
  Hilberath}, and Werth}}]{gerz87}
\bibinfo{author}{\bibnamefont{{Ch. Gerz}}}, \bibinfo{author}{\bibnamefont{{Th.
  Hilberath}}}, \bibnamefont{and}
  \bibinfo{author}{\bibfnamefont{G.}~\bibnamefont{Werth}}, \bibinfo{journal}{Z.
  Phys. D} \textbf{\bibinfo{volume}{5}}, \bibinfo{pages}{97}
  (\bibinfo{year}{1987}).

\bibitem[{\citenamefont{James}(1998)}]{james98a}
\bibinfo{author}{\bibfnamefont{D.~F.~V.} \bibnamefont{James}},
  \bibinfo{journal}{Appl. Phys. B} \textbf{\bibinfo{volume}{66}},
  \bibinfo{pages}{181} (\bibinfo{year}{1998}).

\bibitem[{\citenamefont{Margolis et~al.}(2003)\citenamefont{Margolis, Huang,
  Barwood, Lea, Klein, Rowley, Gill, and Windeler}}]{PhysRevA.67.032501}
\bibinfo{author}{\bibfnamefont{H.~S.} \bibnamefont{Margolis}},
  \bibinfo{author}{\bibfnamefont{G.}~\bibnamefont{Huang}},
  \bibinfo{author}{\bibfnamefont{G.~P.} \bibnamefont{Barwood}},
  \bibinfo{author}{\bibfnamefont{S.~N.} \bibnamefont{Lea}},
  \bibinfo{author}{\bibfnamefont{H.~A.} \bibnamefont{Klein}},
  \bibinfo{author}{\bibfnamefont{W.~R.~C.} \bibnamefont{Rowley}},
  \bibinfo{author}{\bibfnamefont{P.}~\bibnamefont{Gill}}, \bibnamefont{and}
  \bibinfo{author}{\bibfnamefont{R.~S.} \bibnamefont{Windeler}},
  \bibinfo{journal}{Phys. Rev. A} \textbf{\bibinfo{volume}{67}},
  \bibinfo{pages}{032501} (\bibinfo{year}{2003}).

\bibitem[{\citenamefont{Margolis et~al.}(2004)\citenamefont{Margolis, Barwood,
  Huang, Klein, Lea, Szymaniec, and Gill}}]{Margolis04}
\bibinfo{author}{\bibfnamefont{H.~S.} \bibnamefont{Margolis}},
  \bibinfo{author}{\bibfnamefont{G.~P.} \bibnamefont{Barwood}},
  \bibinfo{author}{\bibfnamefont{G.}~\bibnamefont{Huang}},
  \bibinfo{author}{\bibfnamefont{H.~A.} \bibnamefont{Klein}},
  \bibinfo{author}{\bibfnamefont{S.~N.} \bibnamefont{Lea}},
  \bibinfo{author}{\bibfnamefont{K.}~\bibnamefont{Szymaniec}},
  \bibnamefont{and} \bibinfo{author}{\bibfnamefont{P.}~\bibnamefont{Gill}},
  \bibinfo{journal}{Science} \textbf{\bibinfo{volume}{306}},
  \bibinfo{pages}{1355} (\bibinfo{year}{2004}).

\bibitem[{\citenamefont{Kielpinski et~al.}(2000)\citenamefont{Kielpinski, King,
  Myatt, Sackett, Turchette, Itano, Monroe, Wineland, and
  Zurek}}]{PhysRevA.61.032310}
\bibinfo{author}{\bibfnamefont{D.}~\bibnamefont{Kielpinski}},
  \bibinfo{author}{\bibfnamefont{B.~E.} \bibnamefont{King}},
  \bibinfo{author}{\bibfnamefont{C.~J.} \bibnamefont{Myatt}},
  \bibinfo{author}{\bibfnamefont{C.~A.} \bibnamefont{Sackett}},
  \bibinfo{author}{\bibfnamefont{Q.~A.} \bibnamefont{Turchette}},
  \bibinfo{author}{\bibfnamefont{W.~M.} \bibnamefont{Itano}},
  \bibinfo{author}{\bibfnamefont{C.}~\bibnamefont{Monroe}},
  \bibinfo{author}{\bibfnamefont{D.~J.} \bibnamefont{Wineland}},
  \bibnamefont{and} \bibinfo{author}{\bibfnamefont{W.~H.} \bibnamefont{Zurek}},
  \bibinfo{journal}{Phys. Rev. A} \textbf{\bibinfo{volume}{61}},
  \bibinfo{pages}{032310} (\bibinfo{year}{2000}).

\bibitem[{\citenamefont{Rohde et~al.}(2001)\citenamefont{Rohde, Gulde, Roos,
  Barton, Leibfried, Eschner, Schmidt-Kaler, and Blatt}}]{rohde01a}
\bibinfo{author}{\bibfnamefont{H.}~\bibnamefont{Rohde}},
  \bibinfo{author}{\bibfnamefont{S.~T.} \bibnamefont{Gulde}},
  \bibinfo{author}{\bibfnamefont{C.~F.} \bibnamefont{Roos}},
  \bibinfo{author}{\bibfnamefont{P.~A.} \bibnamefont{Barton}},
  \bibinfo{author}{\bibfnamefont{D.}~\bibnamefont{Leibfried}},
  \bibinfo{author}{\bibfnamefont{J.}~\bibnamefont{Eschner}},
  \bibinfo{author}{\bibfnamefont{F.}~\bibnamefont{Schmidt-Kaler}},
  \bibnamefont{and} \bibinfo{author}{\bibfnamefont{R.}~\bibnamefont{Blatt}},
  \bibinfo{journal}{J. Opt. B: Quantum Semiclassical Opt.}
  \textbf{\bibinfo{volume}{3}}, \bibinfo{pages}{S34} (\bibinfo{year}{2001}).

\bibitem[{\citenamefont{Blinov et~al.}(2002)\citenamefont{Blinov, Deslauriers,
  Lee, Madsen, Miller, and Monroe}}]{blinov02a}
\bibinfo{author}{\bibfnamefont{B.~B.} \bibnamefont{Blinov}},
  \bibinfo{author}{\bibfnamefont{L.}~\bibnamefont{Deslauriers}},
  \bibinfo{author}{\bibfnamefont{P.}~\bibnamefont{Lee}},
  \bibinfo{author}{\bibfnamefont{M.~J.} \bibnamefont{Madsen}},
  \bibinfo{author}{\bibfnamefont{R.}~\bibnamefont{Miller}}, \bibnamefont{and}
  \bibinfo{author}{\bibfnamefont{C.}~\bibnamefont{Monroe}},
  \bibinfo{journal}{Phys. Rev. A} \textbf{\bibinfo{volume}{65}},
  \bibinfo{pages}{040304} (\bibinfo{year}{2002}).

\bibitem[{\citenamefont{Barrett et~al.}(2003)\citenamefont{Barrett, DeMarco,
  Schaetz, Meyer, Leibfried, Britton, Chiaverini, Itano, Jelenkovic, Jost
  et~al.}}]{barrett03a}
\bibinfo{author}{\bibfnamefont{M.}~\bibnamefont{Barrett}},
  \bibinfo{author}{\bibfnamefont{B.~L.} \bibnamefont{DeMarco}},
  \bibinfo{author}{\bibfnamefont{T.}~\bibnamefont{Schaetz}},
  \bibinfo{author}{\bibfnamefont{V.}~\bibnamefont{Meyer}},
  \bibinfo{author}{\bibfnamefont{D.}~\bibnamefont{Leibfried}},
  \bibinfo{author}{\bibfnamefont{J.}~\bibnamefont{Britton}},
  \bibinfo{author}{\bibfnamefont{J.}~\bibnamefont{Chiaverini}},
  \bibinfo{author}{\bibfnamefont{W.~M.} \bibnamefont{Itano}},
  \bibinfo{author}{\bibfnamefont{B.~M.} \bibnamefont{Jelenkovic}},
  \bibinfo{author}{\bibfnamefont{J.~D.} \bibnamefont{Jost}},
  \bibnamefont{et~al.}, \bibinfo{journal}{Phys. Rev. A}
  \textbf{\bibinfo{volume}{68}}, \bibinfo{pages}{042302}
  (\bibinfo{year}{2003}).

\bibitem[{\citenamefont{Home et~al.}(2009)\citenamefont{Home, McDonnell, Szwer,
  Keitch, Lucas, Stacey, and Steane}}]{PhysRevA.79.050305}
\bibinfo{author}{\bibfnamefont{J.~P.} \bibnamefont{Home}},
  \bibinfo{author}{\bibfnamefont{M.~J.} \bibnamefont{McDonnell}},
  \bibinfo{author}{\bibfnamefont{D.~J.} \bibnamefont{Szwer}},
  \bibinfo{author}{\bibfnamefont{B.~C.} \bibnamefont{Keitch}},
  \bibinfo{author}{\bibfnamefont{D.~M.} \bibnamefont{Lucas}},
  \bibinfo{author}{\bibfnamefont{D.~N.} \bibnamefont{Stacey}},
  \bibnamefont{and} \bibinfo{author}{\bibfnamefont{A.~M.}
  \bibnamefont{Steane}}, \bibinfo{journal}{Phys. Rev. A}
  \textbf{\bibinfo{volume}{79}}, \bibinfo{pages}{050305}
  (\bibinfo{year}{2009}).

\bibitem[{\citenamefont{{Lybarger, Jr.}}(2010)}]{lybarger10}
\bibinfo{author}{\bibfnamefont{W.~E.} \bibnamefont{{Lybarger, Jr.}}}, Ph.D.
  thesis, \bibinfo{school}{University of California, Los Angeles}
  (\bibinfo{year}{2010}).

\bibitem[{\citenamefont{Berkeland}(2002)}]{berkeland02a}
\bibinfo{author}{\bibfnamefont{D.~J.} \bibnamefont{Berkeland}},
  \bibinfo{journal}{Rev. Sci. Instrum.} \textbf{\bibinfo{volume}{73}},
  \bibinfo{pages}{2856} (\bibinfo{year}{2002}).

\bibitem[{\citenamefont{Vant et~al.}(2006)\citenamefont{Vant, Chiaverini,
  Lybarger, and Berkeland}}]{vant06a}
\bibinfo{author}{\bibfnamefont{K.}~\bibnamefont{Vant}},
  \bibinfo{author}{\bibfnamefont{J.}~\bibnamefont{Chiaverini}},
  \bibinfo{author}{\bibfnamefont{W.}~\bibnamefont{Lybarger}}, \bibnamefont{and}
  \bibinfo{author}{\bibfnamefont{D.~J.} \bibnamefont{Berkeland}},
  \bibinfo{journal}{arXiv:quant-ph/0607055}  (\bibinfo{year}{2006}).

\bibitem[{\citenamefont{Dicke}(1953)}]{PhysRev.89.472}
\bibinfo{author}{\bibfnamefont{R.~H.} \bibnamefont{Dicke}},
  \bibinfo{journal}{Phys. Rev.} \textbf{\bibinfo{volume}{89}},
  \bibinfo{pages}{472} (\bibinfo{year}{1953}).

\bibitem[{\citenamefont{Nagourney et~al.}(1986)\citenamefont{Nagourney,
  Sandberg, and Dehmelt}}]{PhysRevLett.56.2797}
\bibinfo{author}{\bibfnamefont{W.}~\bibnamefont{Nagourney}},
  \bibinfo{author}{\bibfnamefont{J.}~\bibnamefont{Sandberg}}, \bibnamefont{and}
  \bibinfo{author}{\bibfnamefont{H.}~\bibnamefont{Dehmelt}},
  \bibinfo{journal}{Phys. Rev. Lett.} \textbf{\bibinfo{volume}{56}},
  \bibinfo{pages}{2797} (\bibinfo{year}{1986}).

\bibitem[{\citenamefont{Drever et~al.}(1983)\citenamefont{Drever, Hall,
  Kowalski, Hough, Ford, Munley, and Ward}}]{drever83a}
\bibinfo{author}{\bibfnamefont{R.~W.~P.} \bibnamefont{Drever}},
  \bibinfo{author}{\bibfnamefont{J.~L.} \bibnamefont{Hall}},
  \bibinfo{author}{\bibfnamefont{F.~V.} \bibnamefont{Kowalski}},
  \bibinfo{author}{\bibfnamefont{J.}~\bibnamefont{Hough}},
  \bibinfo{author}{\bibfnamefont{G.~M.} \bibnamefont{Ford}},
  \bibinfo{author}{\bibfnamefont{A.~J.} \bibnamefont{Munley}},
  \bibnamefont{and} \bibinfo{author}{\bibfnamefont{H.}~\bibnamefont{Ward}},
  \bibinfo{journal}{Appl. Phys. B} \textbf{\bibinfo{volume}{31}},
  \bibinfo{pages}{97} (\bibinfo{year}{1983}).

\bibitem[{\citenamefont{Myerson et~al.}(2008)\citenamefont{Myerson, Szwer,
  Webster, Allcock, Curtis, Imreh, Sherman, Stacey, Steane, and
  Lucas}}]{myerson08a}
\bibinfo{author}{\bibfnamefont{A.~H.} \bibnamefont{Myerson}},
  \bibinfo{author}{\bibfnamefont{D.~J.} \bibnamefont{Szwer}},
  \bibinfo{author}{\bibfnamefont{S.~C.} \bibnamefont{Webster}},
  \bibinfo{author}{\bibfnamefont{D.~T.~C.} \bibnamefont{Allcock}},
  \bibinfo{author}{\bibfnamefont{M.~J.} \bibnamefont{Curtis}},
  \bibinfo{author}{\bibfnamefont{G.}~\bibnamefont{Imreh}},
  \bibinfo{author}{\bibfnamefont{J.~A.} \bibnamefont{Sherman}},
  \bibinfo{author}{\bibfnamefont{D.~N.} \bibnamefont{Stacey}},
  \bibinfo{author}{\bibfnamefont{A.~M.} \bibnamefont{Steane}},
  \bibnamefont{and} \bibinfo{author}{\bibfnamefont{D.~M.} \bibnamefont{Lucas}},
  \bibinfo{journal}{Phys. Rev. Lett.} \textbf{\bibinfo{volume}{100}},
  \bibinfo{pages}{200502} (\bibinfo{year}{2008}).

\bibitem[{\citenamefont{Woodgate}(1983, c1980)}]{woodgate1983}
\bibinfo{author}{\bibfnamefont{G.~K.} \bibnamefont{Woodgate}},
  \emph{\bibinfo{title}{Elementary Atomic Structure}}
  (\bibinfo{publisher}{Oxford: Clarendon Press}, \bibinfo{year}{1983, c1980}),
  \bibinfo{edition}{2nd} ed.

\bibitem[{\citenamefont{King}({1984})}]{king1984}
\bibinfo{author}{\bibfnamefont{W.}~\bibnamefont{King}},
  \emph{\bibinfo{title}{{Isotope shifts in atomic spectra}}}
  (\bibinfo{publisher}{{Plenum}}, \bibinfo{address}{{New York, NY, USA}},
  \bibinfo{year}{{1984}}), ISBN \bibinfo{isbn}{{0 306 41562 3}}.

\bibitem[{\citenamefont{Heilig and Steudal}(1974)}]{heilig1974}
\bibinfo{author}{\bibfnamefont{K.}~\bibnamefont{Heilig}} \bibnamefont{and}
  \bibinfo{author}{\bibfnamefont{A.}~\bibnamefont{Steudal}},
  \bibinfo{journal}{Atomic Data and Nuclear Data Tables}
  \textbf{\bibinfo{volume}{14}}, \bibinfo{pages}{613} (\bibinfo{year}{1974}).

\bibitem[{\citenamefont{Johnson et~al.}(1988)\citenamefont{Johnson, Blundell,
  and Sapirstein}}]{4johnson88pra}
\bibinfo{author}{\bibfnamefont{W.~R.} \bibnamefont{Johnson}},
  \bibinfo{author}{\bibfnamefont{S.~A.} \bibnamefont{Blundell}},
  \bibnamefont{and}
  \bibinfo{author}{\bibfnamefont{J.}~\bibnamefont{Sapirstein}},
  \bibinfo{journal}{Phys. Rev. A} \textbf{\bibinfo{volume}{37}},
  \bibinfo{pages}{307} (\bibinfo{year}{1988}).

\bibitem[{\citenamefont{Angeli}(2004)}]{5angeli04adndt}
\bibinfo{author}{\bibfnamefont{I.}~\bibnamefont{Angeli}}, \bibinfo{journal}{At.
  Data Nucl. Data Tables} \textbf{\bibinfo{volume}{87}}, \bibinfo{pages}{185}
  (\bibinfo{year}{2004}).

\bibitem[{\citenamefont{Korol and Kozlov}(2007)}]{6korol07pra}
\bibinfo{author}{\bibfnamefont{V.~A.} \bibnamefont{Korol}} \bibnamefont{and}
  \bibinfo{author}{\bibfnamefont{M.~G.} \bibnamefont{Kozlov}},
  \bibinfo{journal}{Phys. Rev. A} \textbf{\bibinfo{volume}{76}},
  \bibinfo{pages}{022103} (\bibinfo{year}{2007}).

\end{thebibliography}

\end{document}